\documentclass[aps,prd,preprint]{revtex4}

\usepackage{longtable}
\usepackage{graphicx}
\usepackage{color}
\usepackage{tikz}
\usepackage{units}
\usetikzlibrary{shapes,arrows}
\usepackage{epstopdf}

\newcommand{\beq}{\begin{equation}}
\newcommand{\eeq}{\end{equation}}
\newcommand{\bdm}{\begin{displaymath}}
\newcommand{\edm}{\end{displaymath}}

\begin{document}

\title{Wiener filtering with a seismic underground array at the Sanford Underground Research Facility}
\author{M Coughlin}
\affiliation{Department of Physics, Harvard University, Cambridge, MA 02138, USA}
\author{J Harms}
\affiliation{INFN, Sezione di Firenze, Firenze 50019, Italy}
\author{N Christensen}
\affiliation{Physics and Astronomy, Carleton College, Northfield, Minnesota 55057, USA}
\author{V Dergachev}
\affiliation{LIGO Laboratory, California Institute of Technology, MS 100-36, Pasadena, CA, 91125, USA}
\author{R DeSalvo}
\affiliation{Collegio di Dottorato, University of Sannio, Benevento 82100, Italy}
\author{S Kandhasamy}
\affiliation{Physics and Astronomy, University of Mississippi, University, MS 38677-1848, USA}
\author{V Mandic}
\affiliation{School of Physics and Astronomy, University of Minnesota, Minneapolis, Minnesota 55455, USA}

\begin{abstract}
A seismic array has been deployed at the Sanford Underground Research Facility in the former Homestake mine, South Dakota, to study the underground seismic environment. This includes exploring the advantages of constructing a third-generation gravitational-wave detector underground. A major noise source for these detectors would be Newtonian noise, which is induced by fluctuations in the local gravitational field. The hope is that a combination of a low-noise seismic environment and coherent noise subtraction using seismometers in the vicinity of the detector could suppress the Newtonian noise to below the projected noise floor for future gravitational-wave detectors. In this paper, certain properties of the Newtonian-noise subtraction problem are studied by applying similar techniques to data of a seismic array. We use Wiener filtering techniques to subtract coherent noise in a seismic array in the frequency band 0.05 -- 1\,Hz. This achieves more than an order of magnitude noise cancellation over a majority of this band. The variation in the Wiener-filter coefficients over the course of the day, including how local activities impact the filter, is analyzed. We also study the variation in coefficients over the course of a month, showing the stability of the filter with time. How varying the filter order affects the subtraction performance is also explored. It is shown that optimizing filter order can significantly improve subtraction of seismic noise.
\end{abstract}


\maketitle

\section{Introduction}
\label{sec:Intro}
In the next few years, a second generation of laser-interferometric gravitational-wave (GW) detectors will start operating with the goal to directly observe GWs. At high frequencies, sources of GWs include core collapse supernovae \cite{OtEA2013,LoOt2012} or the merger of neutron stars and black holes \cite{AbEA2012}. At low frequencies, a stochastic background of GWs, most likely from cosmological origin, is possible \cite{AbEA2009}. The global network of detectors will consist of the two Advanced LIGO \cite{LSC2010} interferometers in Louisiana and Washington state, US, the Advanced Virgo \cite{Vir2011} interferometer near Pisa, Italy, GEO-HF in Hannover, Germany \cite{LuEA2010}, the KAGRA interferometer at the Kamioka mine in Japan \cite{AsEA2013}, and the IndIGO detector in India \cite{Unn2013}. Gravitational waves, which will hopefully be observed by these detectors, produce changes in distance between test masses smaller than $10^{-20}\,$m over kilometer distance scales. Isolating test masses in gravitational-wave detectors from seismic disturbances is one of the foremost challenges of the instrumental design of the detectors. Second-generation detectors will use sophisticated seismic-isolation systems to make possible the detection of GWs above 10\,Hz. For this purpose, the isolation systems consist of chains of coupled springs and pendulums with lowest resonance frequencies around 1\,Hz so that seismic noise well above these frequencies is suppressed by many orders of magnitude in all rotational and translational degrees of freedom. The goal of third-generation detectors, such as the Einstein Telescope \cite{PuEA2010}, will be to extend the detection band to even lower frequencies. This poses an even greater challenge to the design of the isolation system. 

The mechanical, or passive, component of the isolation system is assisted by a complex network of sensors and actuators forming the so-called active seismic isolation \cite{AbEA2004}. Sensor data recorded on some of the mechanical stages of the passive system are used to calculate a feedback force acting on the mechanical stage to suppress seismic disturbances. Another variant of the active isolation scheme is the feed-forward noise cancellation \cite{GiEA2003,DrEA2012,DeEA2012}. For example, data from a seismic sensor deployed directly on the ground close to a test mass can be used to cancel seismic noise further down the isolation chain, provided that there is correlation between motion of the ground and of some part of the isolation system. Feed-forward noise cancellation has also been implemented as part of the interferometer control using light sensors \cite{KoEA2014}.

Seismic disturbances can also affect GW detector test-mass motion via gravitational coupling circumventing the entire seismic isolation system \cite{Sau1984,HuTh1998,Cre2008}. The change in mass density in the rock nearby the detector leads to changes in the local gravitational field, which introduces a force on the test-masses. This so-called Newtonian noise (NN) has never been observed, but it is predicted to be one of the limiting noise sources in second-generation detectors at frequencies between 10\,Hz and 20\,Hz, and for third-generation detectors, such as the Einstein Telescope, down to their lowest frequencies at 2\,Hz. As it is impossible to shield a test mass from NN, another method needs to be found to suppress it. One possibility, at least for future detectors, is to select a site characterized by very low levels of seismic noise \cite{BeEA2010,HaEA2010}. The construction of underground detectors has been proposed for this purpose \cite{PuEA2010}. However, this option is costly, and does not apply to any of the existing surface sites of the LIGO and Virgo detectors. Another idea is to attempt a feed-forward noise cancellation using auxiliary sensors \cite{Cel2000}. For example, gravity perturbations caused by seismic fields can be estimated in real time using data from an array of seismic sensors \cite{DHA2012}.

There are also concepts for a number of future low-frequency GW detectors, called MANGO \cite{HaEA2013}, with sensitivity goals better than $10^{-19} / \sqrt{\rm Hz}$ in the 0.1\,Hz to 10\,Hz band. One class of possible detectors includes atom interferometers, which contain a source of ultracold atoms in free fall that interact multiple times with a laser. Another example is a torsion-bar antenna, which uses tidal-force fluctuations caused by GWs which are observed as differential rotations between two orthogonal bars, independently suspended as torsion pendulums \cite{AnEA2010b,ShEA2014}. A final possibility is using the existing Michelson interferometer detector design optimized to low frequencies. One of the main noise sources in this frequency band will be the NN from seismic surface fields. The study presented in this paper of coherent seismic-noise cancellation in the frequency range 0.05 -- 1\,Hz is a first step to investigate the feasibility of a seismic NN cancellation in the same band. It is therefore directly relevant to maximizing the sensitivity of MANGO detectors.

There are a number of important GW signal sources in the MANGO GW detector band. Compact binaries in their inspiral and merger phase are strong possibilities \cite{HaEA2013}. Intermediate mass black hole binaries can merge in this band \cite{Aa2014}. Galactic white dwarf binaries would likely be detectable \cite{HaEA2013}. Other sources include helioseismic and other pulsation modes \cite{CuLi1996}. Although interesting in their own right, these would be a foreground for the potential detection of primordial GWs. GWs were recently possibly detected in the B-mode polarization of the CMB background, which would provide confirmation of the theory of inflation \cite{AdEA2014}. Assuming a slow roll inflationary model, this signal would correspond to a GW energy density spectrum $\Omega_{\rm GW} \approx 10^{-15}$ in the 0.1\,Hz to 1\,Hz band. Because $\Omega_{\rm GW}(f) \sim S_{\rm GW}(f) f^3$, where $S_{\rm GW}(f)$ is the detector power spectral density, a detector with strain sensitivity of $\sqrt{S_{\rm GW}} \approx 10^{-23} / \sqrt{\mathrm{Hz}}$ in this band might have sufficient sensitivity to detect the inflationary signal. Furthermore, due to the relatively limited astrophysically produced foregrounds, this band appears the most promising for a direct detection of the inflationary signal. 

Newtonian noise directly contributes to the noise of a gravitational-wave interferometer by introducing gravitational forces on the test masses. One of the dominant contributions to Newtonian noise is seismic Newtonian noise. In the following, we will generically refer to seismic Newtonian noise as Newtonian noise. There are two dominant contributions to Newtonian noise: the change in the surface-air boundary caused by seismic waves, and change in the rock densities also caused by seismic waves. There is a linear relationship between the amplitude of seismic waves and Newtonian noise, and thus identification and subtraction of seismic noise in an array of seismometers is useful for potential Newtonian noise subtraction in gravitational-wave interferometers. In the study that follows, we will use an array of seismic sensors to subtract seismic noise from a target seismic sensor, which imitates the time-series of a test mass in a gravitational-wave interferometer.

Therefore, there is significant motivation for exploring techniques which would maximize the sensitivity of detectors in this low-frequency band, and this is the major focus of this paper. For stationary, linear systems, the optimal filter used for a feed-forward noise cancellation is the Wiener filter \cite{Vas2001}. It is calculated from correlations between data of the auxiliary sensors and data observed at the target point where noise is to be suppressed. The parameters of the optimal linear filter under ideal conditions are fully determined by the data correlations and the frequency range over which noise cancellation is to be achieved. In reality though, as will be shown in the following, filter parameters can be further optimized to account for non-stationary properties of the data and variations of the dynamics of the system; one possible example is temperature drift. With respect to slow changes in noise variance or system dynamics, one can simply update the Wiener filter regularly using the latest observed data or implement an adaptive filter technology \cite{Say2003}. However, as will be shown in this paper, in the presence of non-stationary seismic noise, improvement can also be achieved by optimizing the number of filter coefficients. 
 
Wiener filtering with seismic arrays has been performed in the past to improve signal-to-noise ratios towards weak seismic signals \cite{WaEA2008}. In this paper, we present results from a study of feed-forward noise cancellation by means of Wiener filters calculated from correlations between seismometers of an underground array. The sensors are broadband instruments sensitive to seismic noise between about 10\,mHz and 50\,Hz. The array is located at the Sanford Underground Research Facility in the Black Hills of South Dakota \cite{HaEA2010}. The facility has 8 environmentally shielded and isolated stations at 4 different depths. The seismometers are installed on granite tiles placed on concrete platforms connected to the bedrock. They are surrounded by a multi-layer isolation frame of rigid thermal and acoustic insulation panels to further stabilize the thermal environment and to achieve suppression of acoustical signals and air currents. These seismometers have been characterized using huddle tests, and they have been shown to have the same noise floor. Three stations with good data quality were active during this study using data from February and March 2012: one at 800\,ft depth, one at 2000\,ft, and one at 4100\,ft.  Whereas the challenge of seismic Newtonian-noise subtraction cannot be fully represented by our study, it is explained that some key aspects such as stationarity of the noise, and scattering of seismic waves should affect both, Newtonian and seismic-noise subtraction, in similar ways. Therefore, the results of our study allow us to draw certain conclusions for Newtonian-noise subtraction.

In our analysis, the Wiener filters are realized as finite-impulse response (FIR) filters. The two main parameters investigated here are the rate at which the filters are updated, and the number of filter coefficients. A brief summary on data quality issues in GW detectors in general, and specifically of the seismic data used in this study is given in section \ref{sec:DataQuality}. In section \ref{sec:WienerFiltering}, the Wiener filtering method used in this work is described. The results are given in section \ref{sec:results}, and our conclusions are summarized in section \ref{sec:Conclusion}.

\section{Data Quality}
\label{sec:DataQuality}

Despite the fact that the optical system is constructed in vacuum, and the test masses are suspended from seismically isolated platforms, detectors are susceptible to a variety of instrumental and environmental noise sources that decrease their detection sensitivity \cite{ChEA2010,IsEA2010a}. Short in duration, non-astrophysical transient events or \emph{glitches} can mask or mimic real signals. Environmental noise can couple into the interferometer through mechanical vibration or because of magnetic fields which can produce forces on magnets in the suspension systems. Seismic motion from wind, ocean waves, or human activity near the sites are among the most common sources of these disturbances. It is important for filtering methods to be robust against such artifacts, as blindly applying a filter with inputs affected by transients could introduce noise into the target channel. Meadors et al. \cite{MKR2013} successfully applied a subtraction algorithm, based on Allen, Hua, and Ottewill's frequency domain transfer function fitting \cite{AHO1999} to data from LIGO's S6 science run to increase strain sensitivity. To overcome the transients issue, they only applied the algorithm when the subtraction improved the sensitivity. 

In this paper, the reference and target data will come from seismometers of an underground array. Below 1\,Hz, the seismic spectrum is dominated by the primary and secondary microseismic peaks. Far from the ocean, microseisms appear with approximately the same amplitude both at the surface and below ground. These occur between 30 -- 100\,mHz, and 0.1 -- 0.5\,Hz respectively. This is the main source of coherent noise we seek to subtract.  A potential local source of low-frequency seismic disturbances is wind. However, a previous study indicated that at stations at depths of 800\,ft or more, correlation between seismicity and wind speeds is insignificant (while the effect was found to be significant at 300\,ft) \cite{HaEA2010}, implying a relatively short correlation length of the respective seismic disturbance. The strongest sources of low-frequency noise are earthquakes. The closest or highest magnitude events can also significantly contribute to seismic motion above 1\,Hz. Above 1\,Hz, previous work identified the Homestake mine as a world-class low-noise environment \cite{HaEA2010}.

Table~\ref{tab:StationInformation} shows the locations of the seismometers used for our study at the 800\,ft, 2000\,ft-B, and 4100\,ft-A stations.
\begin{table}[ht]
   \begin{center}
   \begin{tabular}{ | l | l | l |}
      \hline
      Station & Seismometer & Position (E,N) (m) \\
      \hline
      800\,ft & T240 & (-88,124)\\
      2000\,ft-B & T240 & (-234,380)\\
      4100\,ft-A & STS-2 & (347,-155)\\
      \hline
   \end{tabular}
   \end{center}
   \caption{Table detailing station information, including station names, type of seismometer, and position relative to Yates shaft of the former Homestake mine. For more details on the mine and available stations, please see \cite{HaEA2010}. T240 stands for Nanometrics Trillium 240 Broadband Sensor, and STS-2 stands for Streckeisen STS-2 Broadband Sensor.}
   \label{tab:StationInformation}
\end{table}
The horizontal distances between them are 800\,{\rm ft}/2000\,{\rm ft-B} = 295\,m, 800\,{\rm ft}/4100\,{\rm ft-A} = 517\,m, and 2000\,{\rm ft-B}/4100\,{\rm ft-A} = 790\,m, and the total distances are 800\,{\rm ft}/2000\,{\rm ft-B} = 503\,m, 800\,{\rm ft}/4100\,{\rm ft-A} = 1236\,m, and 2000\,{\rm ft-B}/4100\,{\rm ft-A} = 1065\,m. The horizontal distances are relevant to estimate coherence from seismic surface fields between stations, whereas the total distances are relevant to estimate coherence from body-wave fields. Whereas the 800\,ft and 2000\,ft-B seismometers provided high-quality data with coherence consistent with the specified instrumental-noise floor of the instruments, 4100\,ft-A showed problems with the data-acquisition system. A frequency comb with 0.5\,Hz spacing was visible in the spectra pointing to a coupling between the timing and readout electronics. Nevertheless, the 4100\,ft-A data were used for this study since the effect on the sub 0.5\,Hz spectrum was a modest increase in the noise floor. It should be emphasized that this excess noise cannot be attributed to the seismometer itself since previous self-noise measurements yielded similar noise levels for all instruments.

Horizontal and vertical channels have very similar spectra at underground sites of the Sanford Underground Laboratory \cite{HaEA2010}. Nevertheless, in the following, the study of Wiener filters and feed-forward noise cancellation will only make use of the vertical seismic channels. Vertical seismic displacement is what produces the by far strongest contribution to NN in MANGO type detectors \cite{HaEA2013}. Horizontal channels are more relevant to seismic isolation in this frequency band, and the additional challenge of using horizontal channels would be to disentangle tilt motion from true horizontal motion, which makes it difficult to compare measured coherence and Wiener-filter performance with theoretical models. In the future, use of a tiltmeter along with the seismometers could be used to disentangle the effect \cite{DeDe2014}.

\section{Wiener Filtering}
\label{sec:WienerFiltering}

%In the following, we will describe feed-forward noise cancellation using seismic data. 
When creating a feed-forward filter to subtract seismically induced noise in GW detectors, the target point can either be located within the seismic isolation chain monitored by a seismic sensor, or perhaps more importantly, it can be the test mass or other optics whose positions are read out interferometrically. The reference sensors used as filter inputs typically include seismometers surrounding the respective optics \cite{DrEA2012,DHA2012}. Therefore, while the first use of these filters has been to supplement the performance of active seismic attenuation, in the more important application of Wiener filtering to GW detectors, the seismometers on site can be used to subtract noise from the detector output. Since there are in general several input sensors and one target point, which is the differential motion between two mirrors, the filter configuration is also known as multiple-input, single output (MISO). The goal is to design a MISO filter that combines the data from all seismometers to produce an optimal estimate of the motion of the target point. This allows removal of seismic noise coupled to the detector data, including NN, without a priori assumptions about the modal content and directionality of the seismic waves. Instead, the filter is derived solely from the input and output data via the filter optimization process.

The type of MISO filter produced depends on the type of noise. If the noise is modelled as a statistically stationary process (one whose parameters are constant in time), the optimal filter is constant in time. Then, the filter is only created once and thereafter can be applied to the data continuously. These types of filters satisfy the Wiener-Hopf equations and are known as \emph{Wiener filters} \cite{Vas2001}. The simplest kind of Wiener filter to implement is a causal FIR Wiener filter, which is characterized by a vector of real numbers that represent an impulse response. This vector is convolved with the input signal in the time domain to obtain the optimal estimate of the noise.

The basic principle of Wiener filters is to use the output of a number of reference traces to predict the noise of a target channel. This prediction is then subtracted from the target channel, leaving the non-correlated residual. While using Wiener filters to supplement seismic attenuation must be done in real time, NN noise removal from the detector output can be done later, which potentially allows for filter optimization. Given the input time series $\vec x(k)=(x_m(k))$ with $m=1,\ldots,M$ from $M$ reference channels, and output time series $y(k)$ from a single target channel, we desire to create an FIR filter with $L+1$ coefficients $\vec f(l)=(f_m(l)),\,l=0,\ldots,L$ that minimizes the residual error:
\begin{equation}
E = \sum_{k=1}^N \left(y(k)-\sum_{l=0}^L \vec f(l)\cdot\vec x(k-l)\right)^2
\end{equation}
where $N$ is the number of samples. Each coefficient of the optimal filter must obey
\begin{equation}
\frac{\delta E}{\delta f_m(l)}  = 0
\end{equation}
This constraint results in
\begin{equation}
\sum_{l=0}^L R_{x x} (k-l)\cdot\vec f(l)  = \vec c_{y x} (k)
\label{eq:filtercoeff}
\end{equation}
where $R_{x x}$ is the autocorrelation of $\vec x$, $\vec c_{y x}$ is the cross-correlation between $y$ and $\vec x$, and $k\in 0,\ldots,L$. Equation (\ref{eq:filtercoeff}) is known as the Wiener-Hopf equation, which represents a series of $(L+1)M$ equations for the same number of optimal filter coefficients $f_m(l)$ that must be solved simultaneously. Because the signals are real, their autocorrelation is symmetric, $R_{x x}(k)=R_{x x}(-k)$, resulting in a Toeplitz structure for the system of equations, which makes it possible to apply more efficient algorithms when solving for the filter coefficients. This solution returns the $f_m$ terms. The noise cancellation algorithm can then be written symbolically as a convolution (symbol $*$) \cite{Vas2001}:
\begin{equation}
r(k) = y (k)-\sum\limits_{m=1}^M (f_m*x_m)(k)
\end{equation}
Because the noise-cancellation filter performs best during times without major seismic disturbances such as earthquakes, one needs to make sure that the data used to calculate the filter coefficients are representative of quiet times. Equation (\ref{eq:filtercoeff}) shows that the optimal filter coefficient depends on the average correlation between channels, but also on the spectra of all channels (otherwise optimal subtraction could be achieved with a single FIR filter coefficient per input channel). Filter properties can be related to specific properties of the seismic wave during strong ground motion, such as propagation direction, or wave content (surface waves, body waves, polarization, etc). The spectrum of a seismic disturbance is typically determined by its source. The typical situation, however, will be that the seismic field recorded over some time is composed of many seismic waves from different sources, and the Wiener filter calculated from it will be determined by the average correlations between channels and their average spectra.

It should be noted that coherence between channels needs to be very high even for ``modest'' noise cancellation. The ideal suppression factor $r(f)$ as a function of frequency $f$ in the case of a single reference channel is related to the reference-target coherence $c(f)$ via
\beq
r(f)=\frac{1}{\sqrt{1-c(f)^2}}
\eeq
where
\begin{equation}
c(f) = \frac{\overline{\tilde{s_1}(f) \tilde{s_2}(f)^*}}{\overline{|\tilde{s_1}(f)|} \overline{|\tilde{s_2}(f)}|}
\end{equation}
and $\tilde{s_1}(f)$ and $\tilde{s_2}(f)$ are the Fourier transforms of the two channels. For example, if coherence between reference and target channels at some frequency is (0.9, 0.99, 0.999), then the residual amplitude spectrum at that frequency will ideally be reduced by factors (2.3, 7.1, 22) respectively. Therefore, a seismic noise reduction by a factor 50 (as achieved here with two reference channels; see section \ref{sec:OptimalWienerFiltering}) is very high corresponding to a channel coherence of about 0.9998. We can compare this result with a theoretical prediction from a Rayleigh-wave model. If only one reference channel were used, i.~e.~taking the 800\,ft seismometer as target and the 2000\,ft-B as reference, and the Rayleigh-wave field were isotropic, then the channel coherence would be about 0.999 assuming a Rayleigh-wave speed of 3\,km/s and using equation (7) in \cite{DHA2012}. This means that a factor 50 subtraction is better than what can ideally be achieved with two channels (under the assumption that the Rayleigh field is isotropic). It can be concluded that the array of two reference channels already provides some ability to measure propagation directions of Rayleigh waves, or to disentangle different wave types and thereby significantly improving the subtraction performance (again, assuming an isotropic Rayleigh field).

\section{Filter Results}
\label{sec:results}
Coherence between channels is what fully determines the Wiener filter coefficients. Coherence is a consequence of the seismometer self noise and the quality of its connection with the ground, but also of the local geophysical settings and properties of the seismic field. Whereas the hard rock of the Black Hills should be beneficial to seismic coherence between stations as seismic waves are relatively long, topographic scattering, especially in mountainous regions, may pose an ultimate limit to what can be achieved with Wiener filtering irrespective of the sensor self noise \cite{CoHa2012}. Another limit of subtraction performance could be related to a complex composition of the seismic field. With the few seismometers that were used in this study, the Wiener filter is likely not able to subtract ambient noise from body waves to the same degree as the noise from the dominant Rayleigh waves. The latter is not obvious though, and needs to be investigated in the future using extended underground arrays. The basic idea is that a Wiener filter can only train on an average correlation pattern of a seismic array. 
\begin{figure}[ht!]
 \includegraphics[width=3.3in]{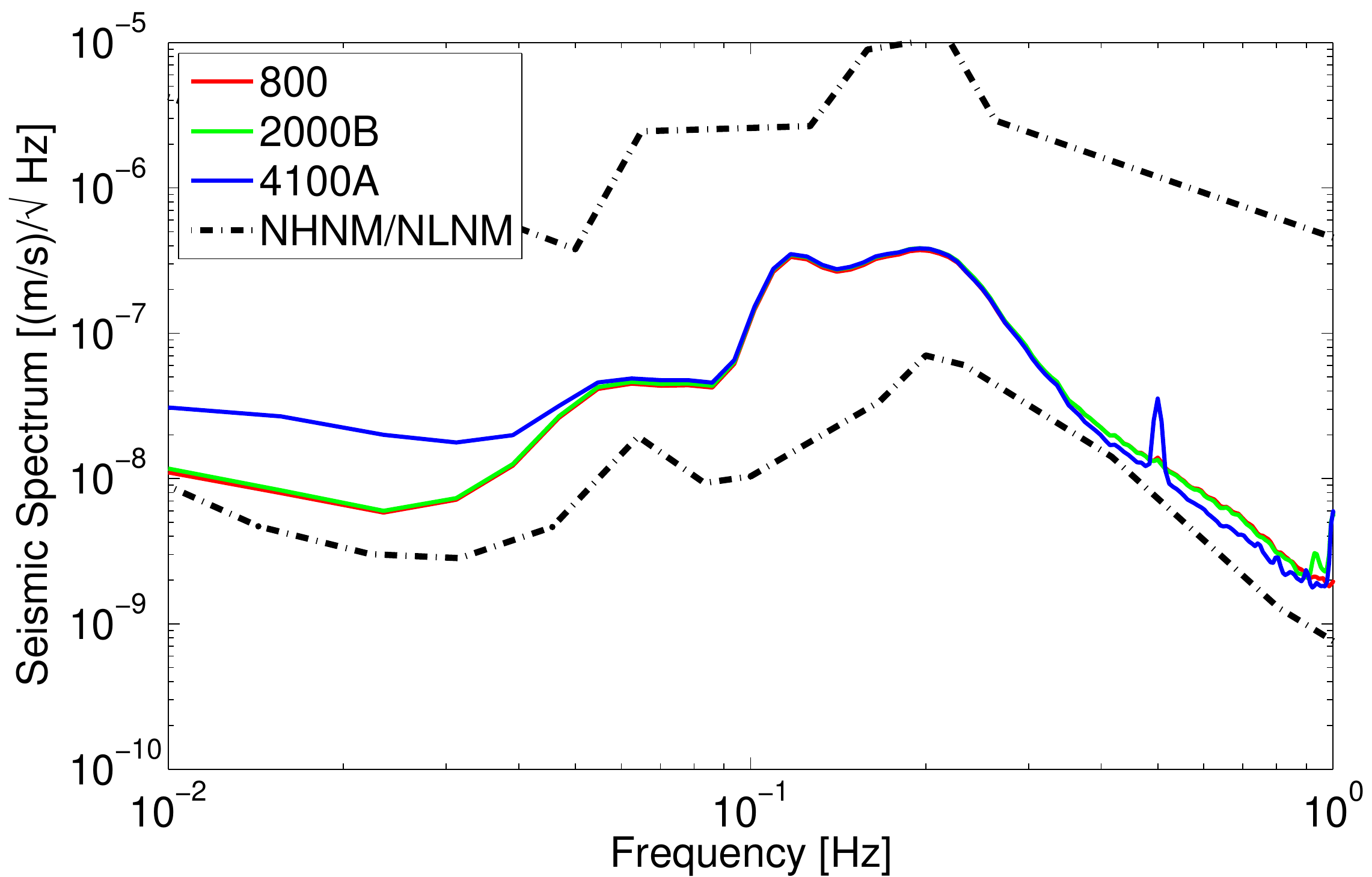}
 \includegraphics[width=3.3in]{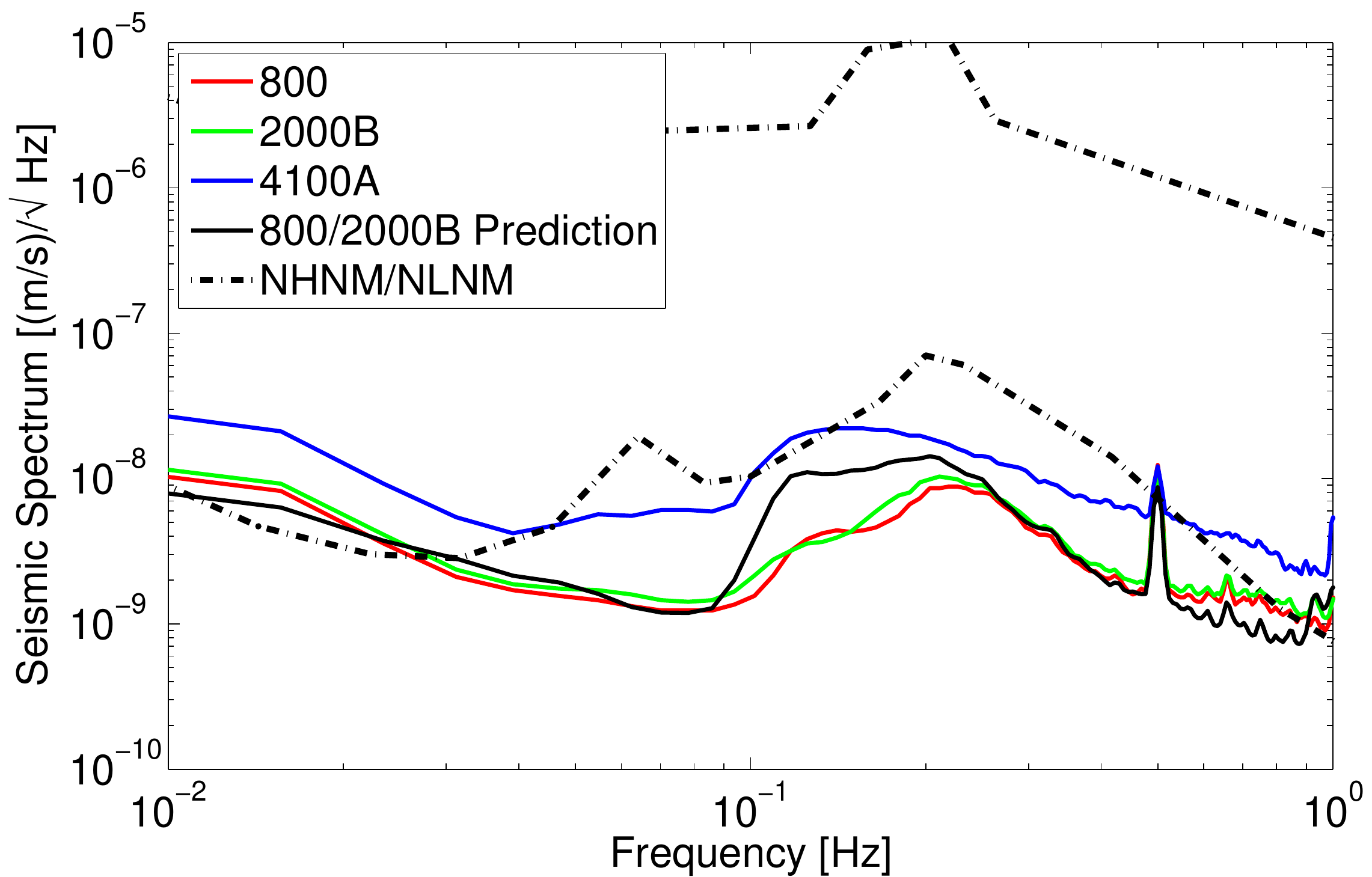}
 \caption{The plot to the left shows the medians of seismic spectra for the three seismometers. The dash-dotted lines in black represent the global new low- and high-noise models (NLNM/NHNM) \cite{Pet1993}. The primary and secondary microseismic peaks are visible between 30 and 100\,mHz and 0.1-0.5\,Hz respectively. A line at 0.5\,Hz from the data-acquisition is also visible. On the right is the median of the residuals for the three seismometers. The vertical channel of the respective seismometer was the target channel, and channels from the other two sensors were used as reference channels. The expected subtraction for the 800\,ft channel based on coherence between the 800\,ft and 2000\,ft-B stations is plotted in solid black.}
 \label{fig:WienerPerformance}
\end{figure}
The sign of average seismic correlation between two seismometers can be different for body and surface waves depending on the distance between them. This leads to partial cancellation of their contributions, which affects subtraction performance. However, if a seismic array is designed to have the ability to distinguish between body and surface fields, then also the Wiener filter will be able to subtract contributions from both.

Figure \ref{fig:WienerPerformance} demonstrates the performance of the filter on the seismic array data. A Wiener filter with 1000 coefficients was calculated from 3 hours of data, and then subsequently applied to subtract ambient seismic noise from target channels over a period of 2 weeks. In all cases, only the vertical channels were used as target and reference. Adding horizontal channels to the reference channels does not lead to significant changes of the subtraction residuals since coherence between horizontal and vertical directions is small. Due to similar coherence between horizontal channels, we expect the subtraction results for vertical channels to hold for horizontal channels as well. For target seismometers 800\,ft and 2000\,ft-B, subtraction residuals are more than a factor 10 weaker than the original spectra over a wide range of frequencies. Subtraction performance is significantly worse at 4100\,ft-A. The reason for this is the elevated noise floor of the 4100\,ft-A seismometer, because of problems with the data-acquisition system as explained in Section \ref{sec:DataQuality}. Nevertheless, it is interesting that in all cases a residual microseismic peak (probably two peaks generated by two different ocean-wave fields) remains in the residual spectrum, and that this peak is larger than what is expected from an estimate of the Rayleigh wavelength (more than 10\,km) in comparison to the horizontal station distance. This can be due to a geophysical effect, but potentially also because of a weakly non-linear response of the seismometer. 

As shown in figure \ref{fig:WienerPerformance}, the subtraction residuals are greater than the coherence limit above 0.5 Hz, which means that the Wiener filter injects uncorrelated noise into the target channel. Ideally, this should be impossible, since injection of noise by Wiener filters is suppressed consistent with the loss of coherence due to this noise (e.~g.~if reference and target channels are uncorrelated, then the Wiener filter function is equal to 0). Therefore, the fact that noise is increased can either be explained by numerical problems in the calculation of the Wiener filter (which involves the inversion of a large matrix), or potentially with non-stationarity of the noise (which can make the Wiener filter sub-optimal). Additional filters such as low-pass filters can be applied in sequence with the Wiener filter to suppress excess noise at certain frequencies.

\begin{figure}[ht!]
 \centerline{\includegraphics[width=5in]{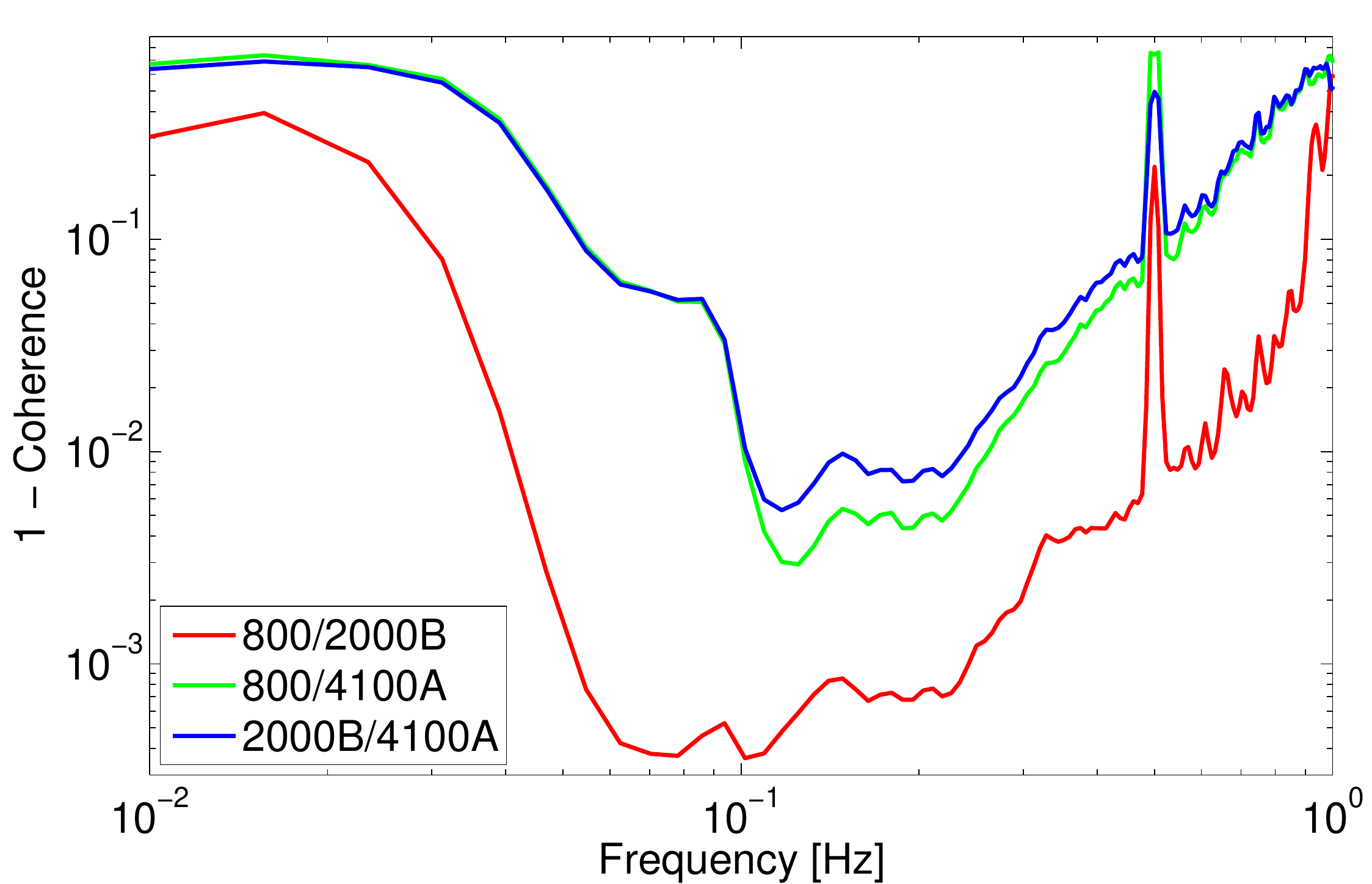}}
 \caption{Plot of of the measured coherence between the three station pairs. Coherence between the 800\,ft and 2000\,ft-B stations exceeds 0.9995 at the microseism. Coherence between the other pairs is approximately an order of magnitude lower, which results in the relatively worse subtraction results presented in the paper.}
 \label{fig:Coherence}
\end{figure}

Figure~\ref{fig:Coherence} presents the coherence between the stations. The measured coherence between two stations can be used to calculate the spectrum of the uncorrelated seismic noise. The result is shown as black solid line on the right of figure \ref{fig:WienerPerformance} for the 800\,ft -- 2000\,ft-B seismometer pair. As can be seen, the residual spectrum of the Wiener filter lies below the coherence limit, which means that with one additional reference seismometer, significant extra information about the seismic field can be obtained to improve noise cancellation. 

\subsection{Filter evolution with time}

Slow changes of average properties of the seismic field (e.~g.~diurnal or seasonal cycles), of instrumental noise, or of the system dynamics may require filter adaptation. When implementing a Wiener filter, it is important to know how often such a filter should be updated, or in case that the required update rate is high, if a genuine adaptive filter technology needs to be implemented. We now study how the filter coefficients evolve over time by recalculating the Wiener filter regularly. 

\begin{figure}[ht]
 \centerline{\includegraphics[width=3.2in]{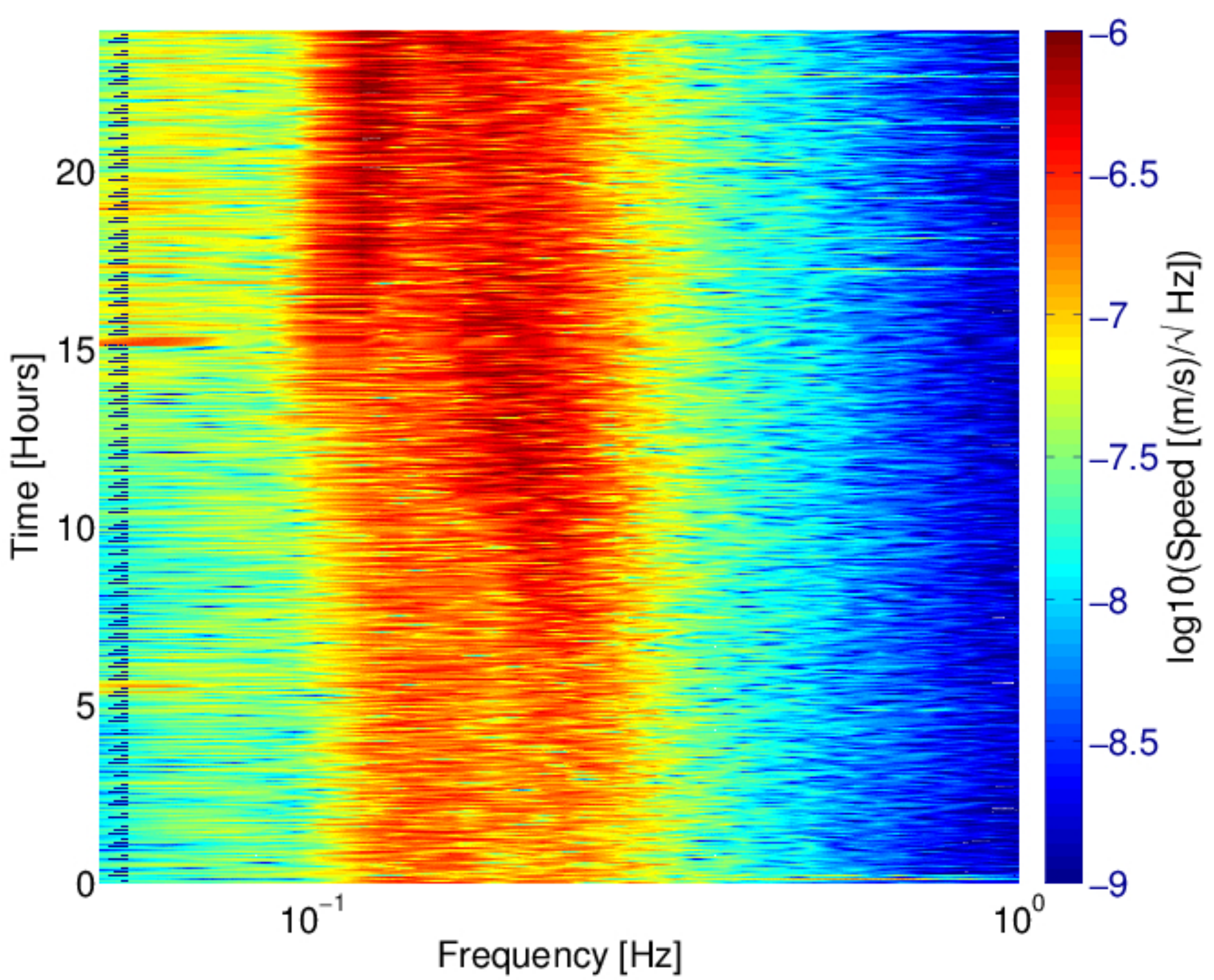}
 \includegraphics[width=2.9in]{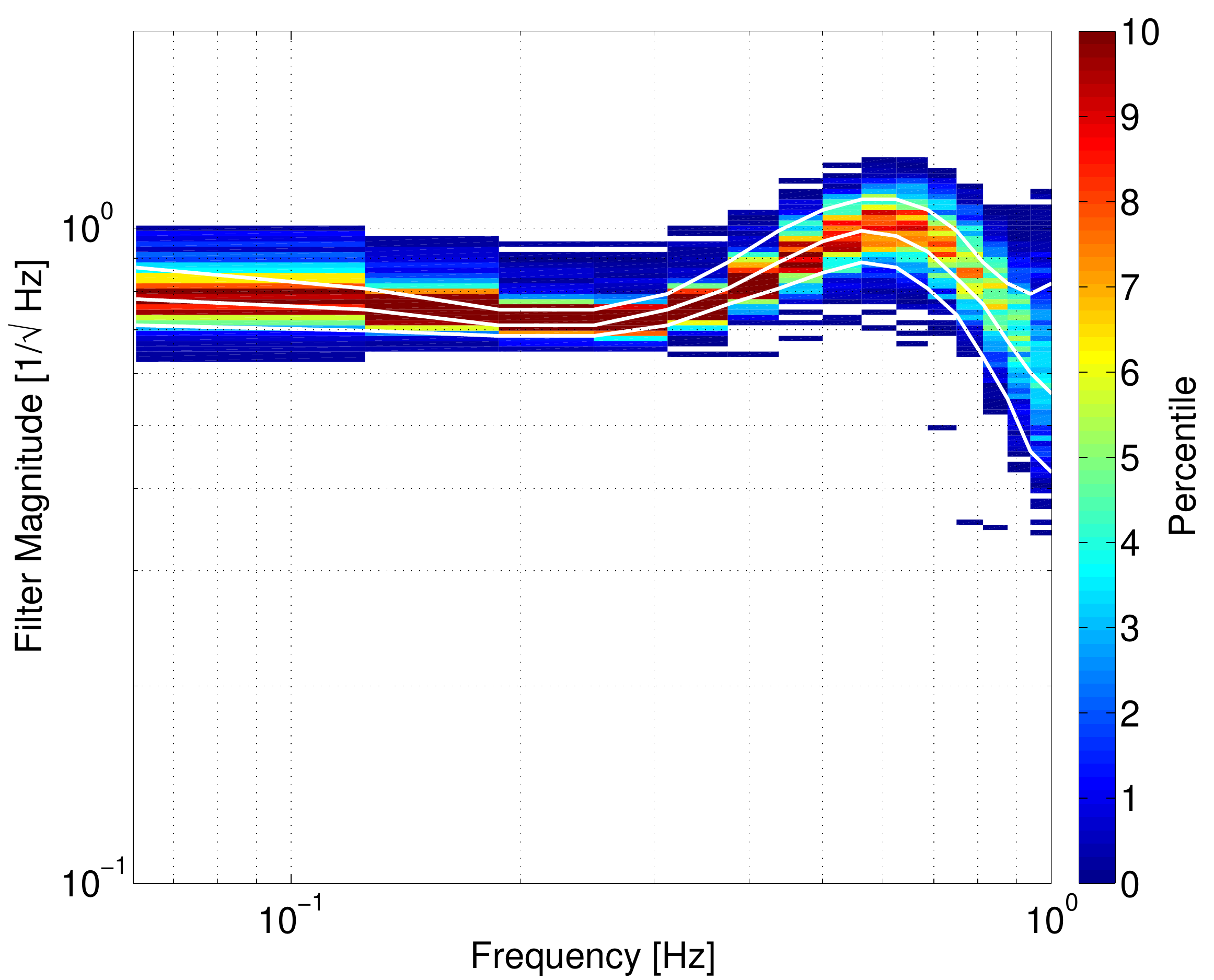}}
\centerline{\includegraphics[width=3.1in]{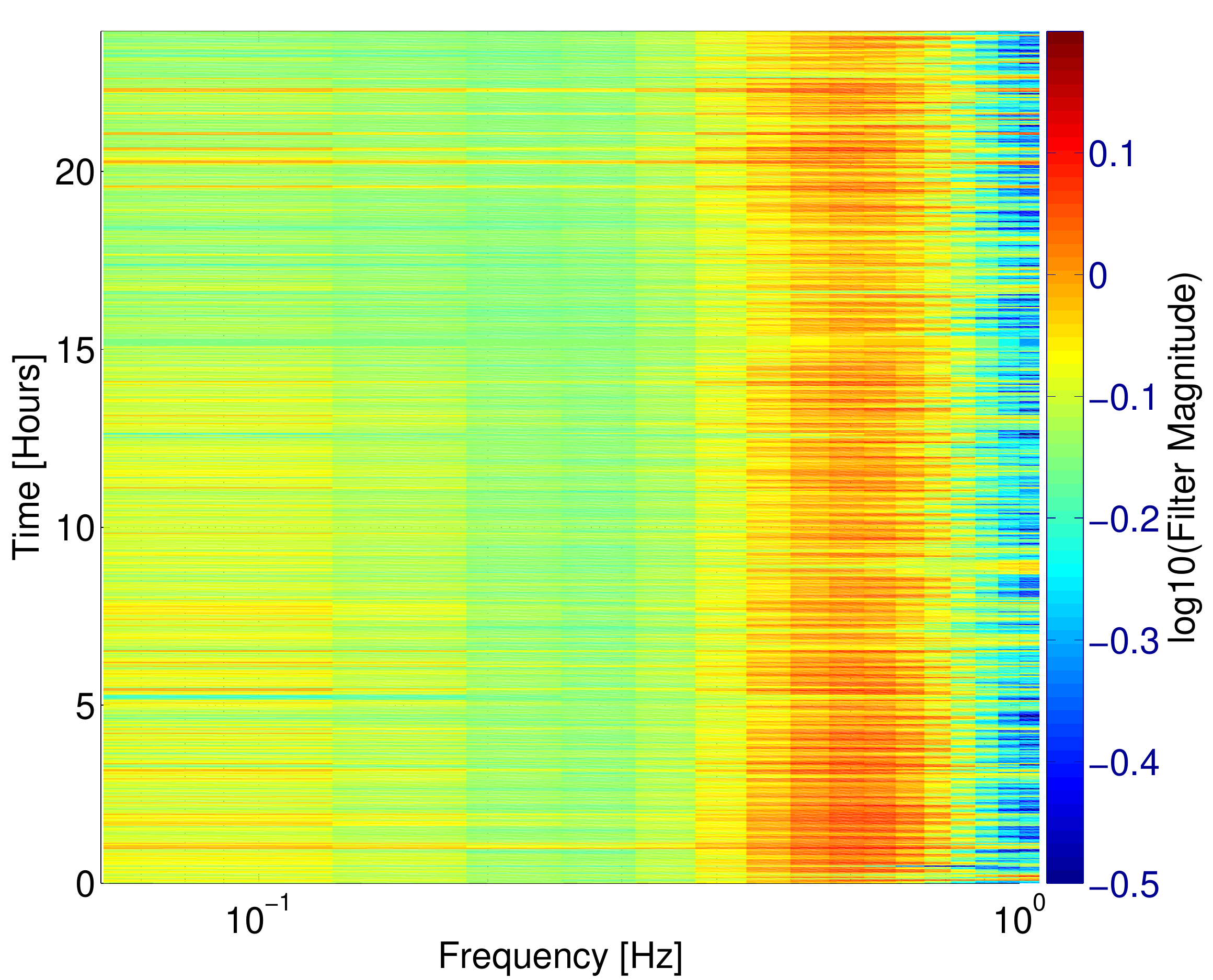}
 \includegraphics[width=3.1in]{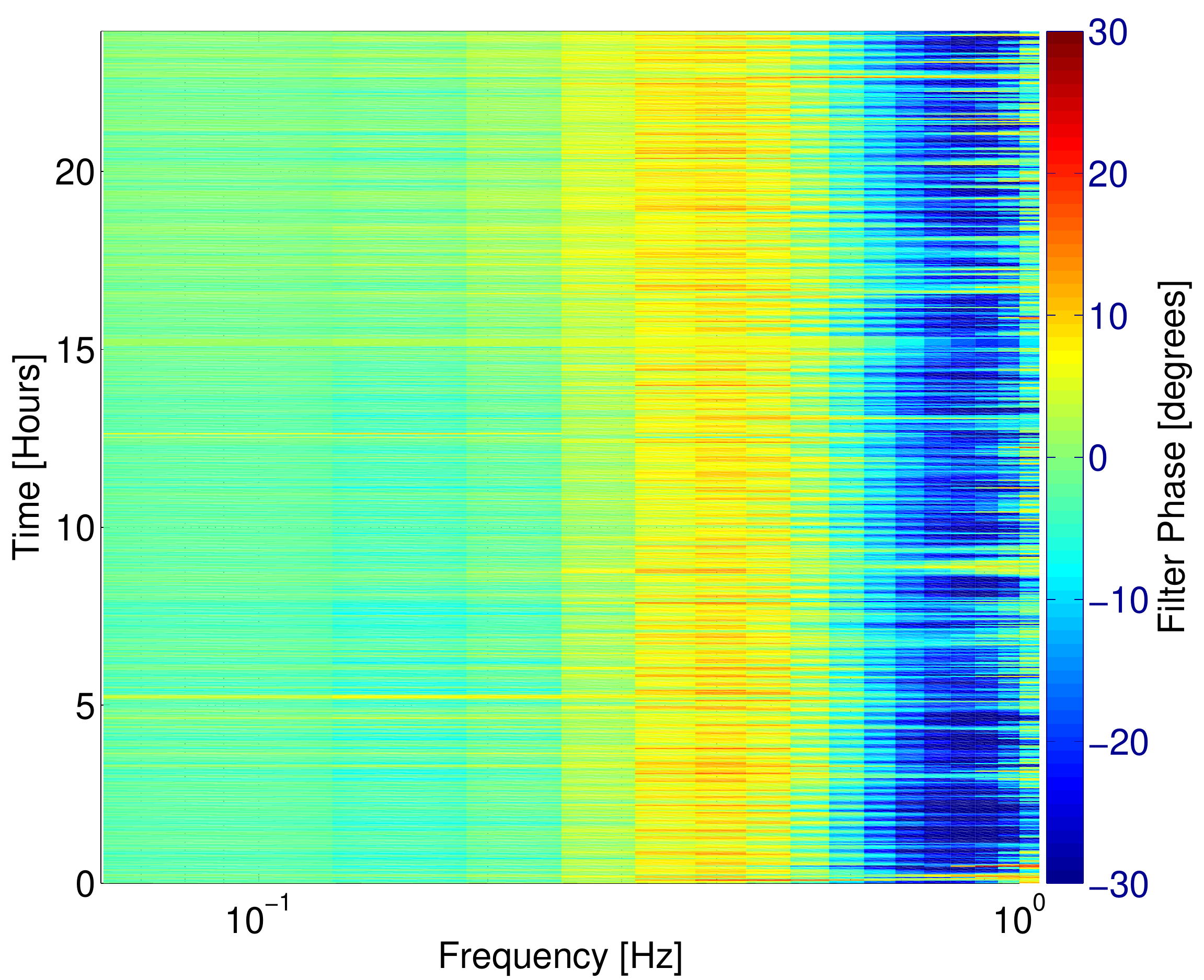}}
 \caption{Filter variation over the course of a day. The filter is represented in the frequency domain by its magnitude and phase, as in a Bode plot. Top left: Time-frequency plot of the spectra of the 800\,ft seismometer. Top right: Filter magnitude histogram for the 2000\,ft-B reference channel. The bottom, middle, and top white lines correspond to the 10th, 50th, and 90th percentiles respectively. Bottom left: Time-frequency plot of the filter magnitude for the 2000\,ft-B reference channel. Bottom right: Time-frequency plot of the filter phase for the 2000\,ft-B reference channel.}.
 \label{fig:FIRspecvarDay}
\end{figure}

We begin by examining filter evolution on day-long timescales by computing a filter every 128\,s. An update scheme this fast is certainly not meant to be used in real applications and only serves to illustrate filter variations. The target seismometer is at the 800\,ft station. Figure \ref{fig:FIRspecvarDay} shows the evolution of the Wiener filter over the course of a day. Results are given only for the part of the filter mapping the 2000\,ft-B reference channel. Instead of directly showing the FIR coefficients, the filter is represented in the frequency domain to make it easier to connect its variation with properties of the seismic field.

The fluctuations in the seismic spectra predominantly result in changes in the high frequency portion of the filter. The major variations coincide with high amplitude events in the seismic spectra. Due to its high frequency, this noise is likely local and therefore anthropogenic in origin. The histogram of filter magnitude variations is shown in the top right plot. The distribution widens at high frequency, while it is narrowest around the oceanic microseisms. This is consistent with the fact that oceanic microseisms are stationary over the course of a day, whereas local noise below and above the microseismic peak is strongly non-stationary. It is maybe surprising that the filter below 0.4\,Hz does not vary as much as the seismic field itself, but the variation in the filter is determined by variations of correlation between all channels. In the case of high correlation between channels during quiet times, as is the case at lower frequencies, the effect of strong ground motion on correlation would simply be to increase correlation from say 0.9 to 0.95, which means that also the filter response should be expected to vary accordingly. Its efficacy would also increase. Correlation is, however, small at higher frequencies during quiet times, and is increased significantly by strong ground motion, which would lead to large changes in the Wiener filter. 

Because the spectral densities of all three channels involved in the noise subtraction are almost identical in the frequency range 0.1\,Hz -- 0.5\,Hz, the fact that filter magnitudes for the 2000\,ft-B are smaller than 1 in this frequency range means that the 4100\,ft-A reference channel provides the missing part required for subtraction. The 4100\,ft-A contribution is smaller due to the elevated noise floor of the seismometer because of data-acquisition problems. The filter magnitudes greater than 1 around 0.8\,Hz are also expected because the target channel is closer to the surface and therefore has a significantly greater Rayleigh-wave amplitude at sufficiently high frequencies. The filter magnitude goes to zero at frequencies where coherence is weak, and the optimal filter suppresses the incoherent noise contribution of the reference channel to the target channel.

\begin{figure}[ht]
\centerline{ \includegraphics[width=5in]{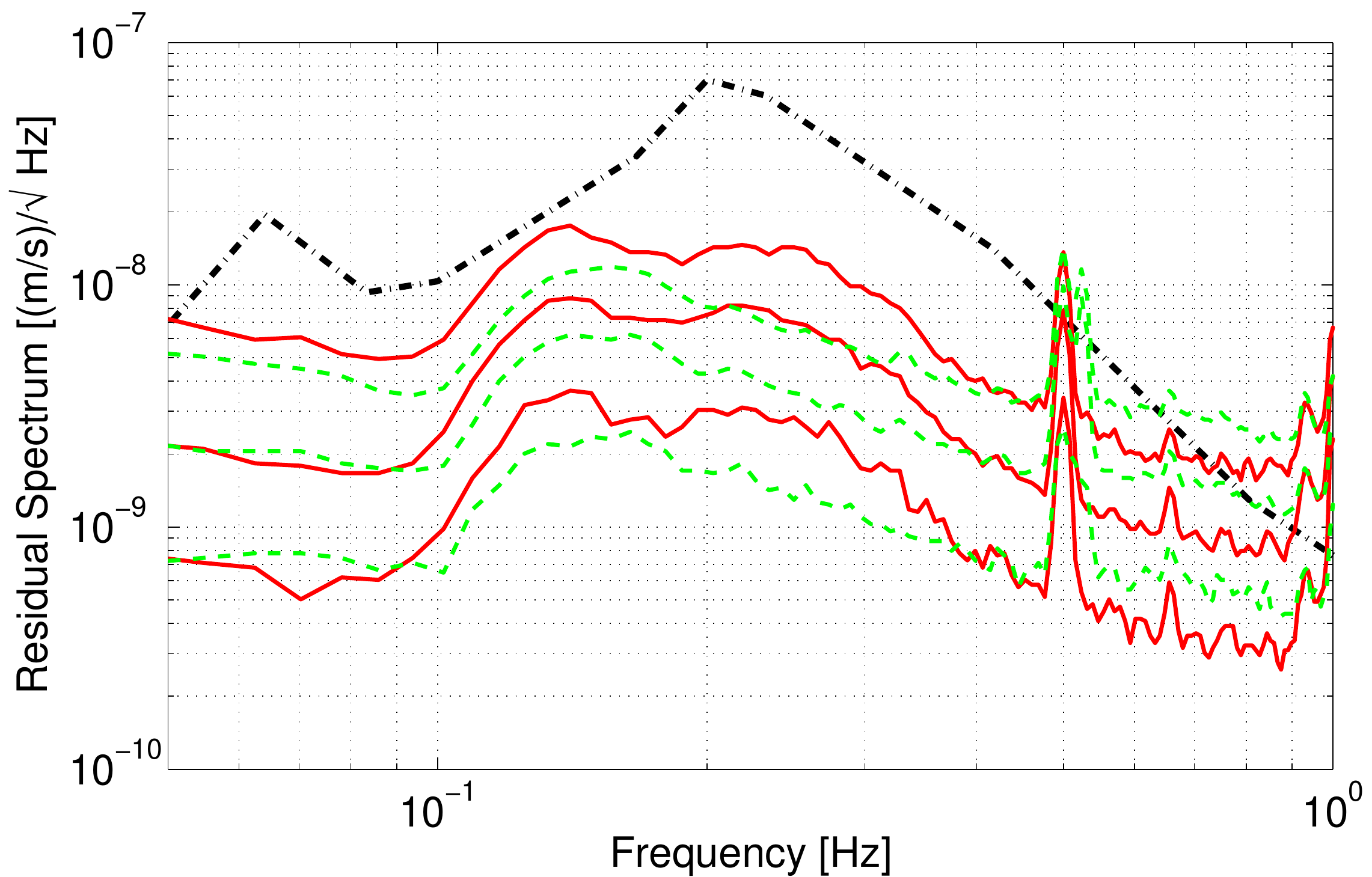}}
 \caption{Subtraction residuals for the 800\,ft channel using 2000\,ft-B and 4100\,ft-A as reference channels over a month of data. The dashed curves are its 10th, 50th, and 90th percentiles using a filter that is updated once every day. The solid lines are the percentiles of residual noise using a filter updated once every week. The dash-dotted line in black represents the global new low-noise model \cite{Pet1993}.}.
 \label{fig:FIRtime}
\end{figure}

Next we study how the filter evolves on month-long timescales. Figure \ref{fig:FIRtime} shows the 10th, 50th, and 90th percentiles of residual noise of the 4100\,ft-A station accumulated over the course of two months. The filter is either updated once a day, leading to the dashed-line percentiles, or once per week leading to the solid-line percentiles. The filter order was 1000 calculated from 2 hours of data.
The subtraction performance is almost identical in the two cases, which means that seismic correlations averaged over one day do not change significantly within one week. It is in fact remarkable that the residual spectra are so similar over a wide range of frequencies.

\subsection{Optimal filter order}
\label{sec:OptimalWienerFiltering}

When designing and optimizing a Wiener filter, there are only a few tunable parameters. This includes the number of reference channels that one uses to create the filter, as well as what kind of additional filtering (if any) to apply to the input data. Another is the order of the Wiener FIR filter. Implementations in GW interferometers \cite{GiEA2003,DrEA2012,DeEA2012} have typically used thousands of filter coefficients at 64\,Hz sampling rate. These filters were calculated using 1 hour of data.

One may wonder whether the order of the most efficient Wiener filter, i.~e.~the filter with minimal order achieving the subtraction goal, only depends on the bandwidth of the noise that one wants to subtract. This assumption seems reasonable, because the filter order and time interval between two filter coefficients determines the minimum and maximum frequency of the filter response. It will be shown in this section that the dependence of subtraction residuals on filter order is not so simple.

\begin{figure}[ht]
 \centerline{\includegraphics[width=3.1in]{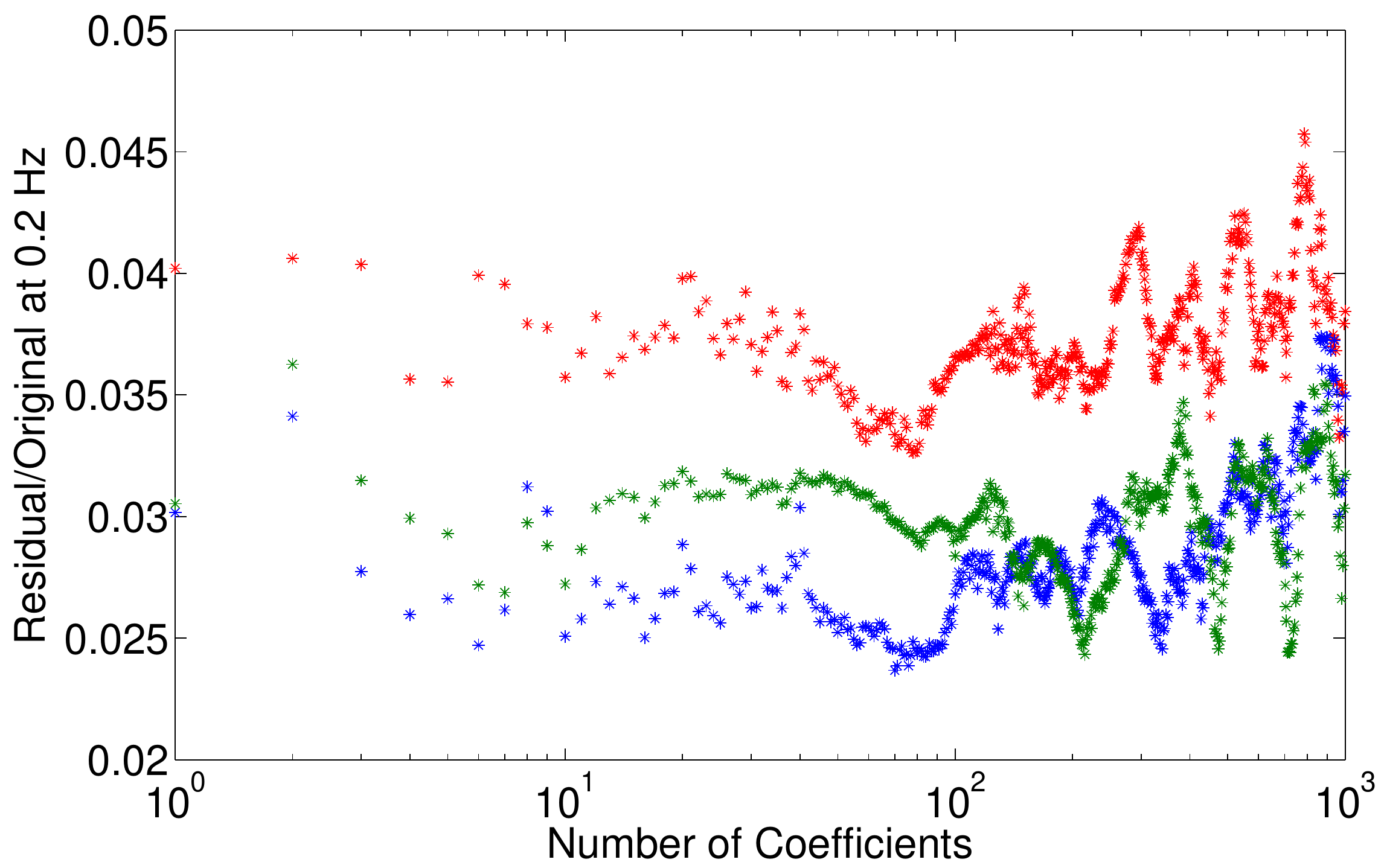} 
 \includegraphics[width=3.1in]{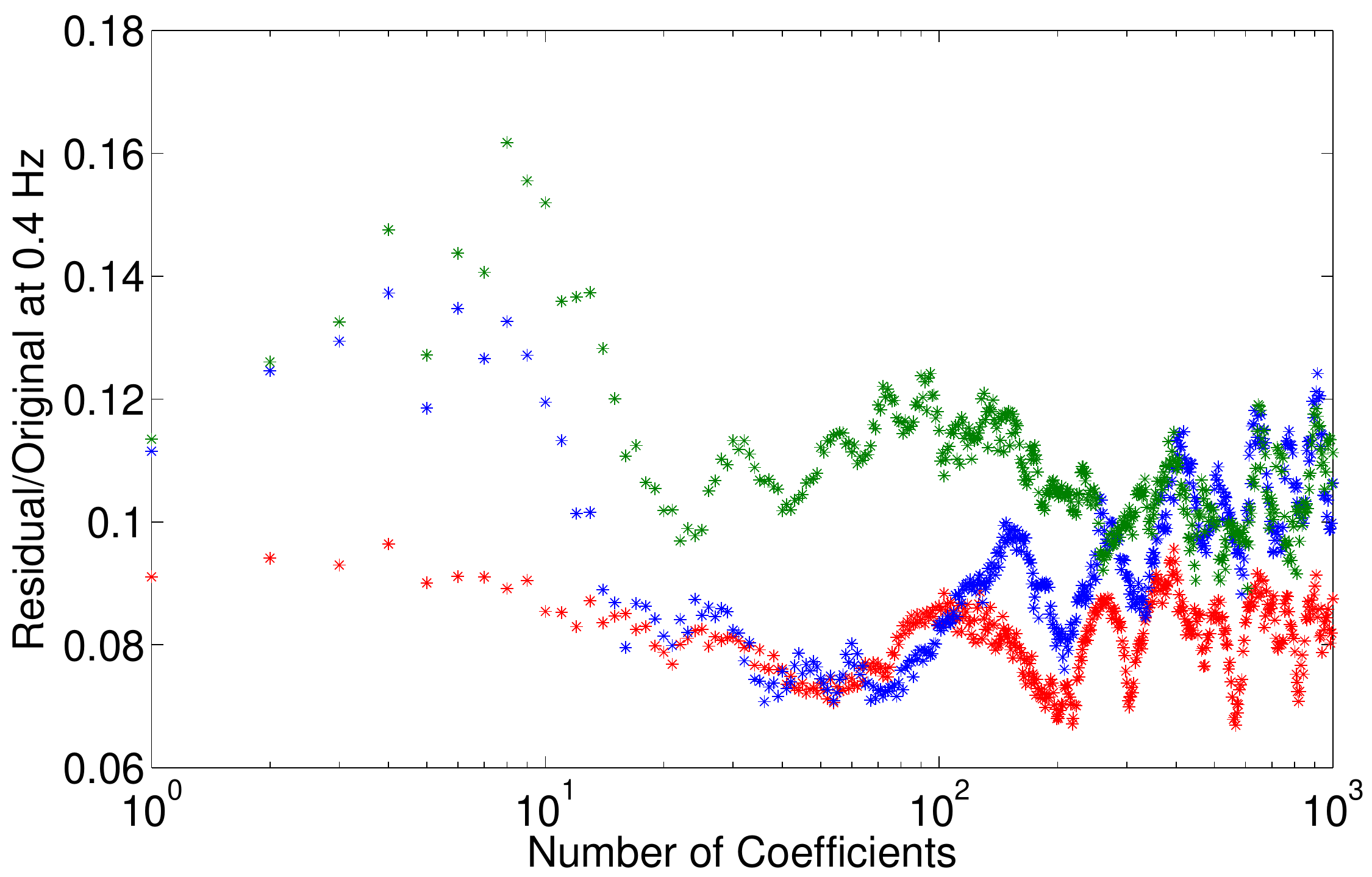}}
 \caption{Residual noise evaluated over the three separate days as a function of filter order for the 800\,ft target channel. The residuals shown in the plot on the left are evaluated at 0.2\,Hz, whereas the residuals on the right are evaluated at 0.4\,Hz. Each color corresponds to a different day.}
 \label{fig:FIROrder}
\end{figure}

We examine the effect of filter order using three separate days of data. For each day, filters are calculated with varying filter order. In figure \ref{fig:FIROrder}, the performances of the filters are compared. The plot on the left shows the noise reduction at 0.2\,Hz (arbitrarily chosen) for the three days and filter orders 1 to 1000, while the plot on the right is the same for 0.4\,Hz. Filter order 1 means simple subtraction of data with optimal scaling factor of the input data.
Subtraction performance shows less variation from one day to the next for high filter orders, but the price paid for this stability is that the residuals could be much less at certain days using a smaller filter order. In general, the best subtraction performance is only achieved for specific filter orders, often degraded substantially when changing from optimal filter order to slightly higher or smaller orders. In addition, the optimal filter order is not the same every day, although it could be optimized for each individual day. However, the plots also show that even though choosing the optimal filter order depends on target frequency and varies between days, it is still possible to identify ranges of filter order that work better than others. For example, for the two frequencies 0.2\,Hz and 0.4\,Hz, figure \ref{fig:FIROrder} shows that a filter order slightly below 100 would be a better choice instead of a very high filter order around 1000.
\begin{figure}[ht]
 \centerline{\includegraphics[width=5in]{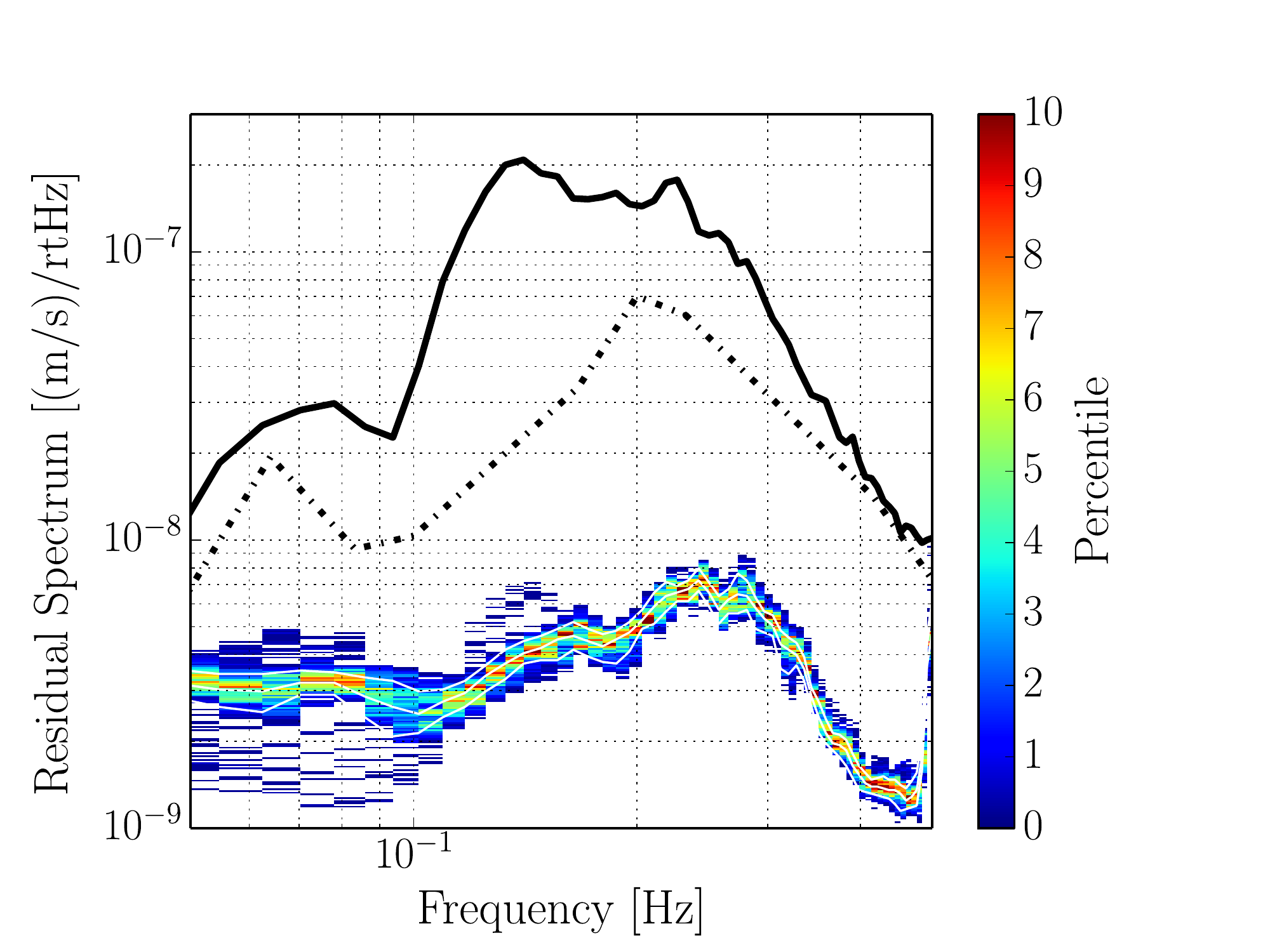}}
 \caption{The plot shows the histogram of residual spectra for the 800\,ft target channel achieved by each of the 1000 filter orders for the day represented by the blue markers in figure \ref{fig:FIROrder}. The bottom, middle, and top white lines correspond to the 10th, 50th, and 90th percentiles respectively. The dash-dotted line in black shows the global new low-noise model \cite{Pet1993}. The original spectra of the 800\,ft channel is plotted in solid black.}
\label{fig:orderhist}
\end{figure}
A full histogram of residuals achieved for the day represented by the blue markers in figure \ref{fig:FIROrder} accumulated from all filter orders between 1 and 1000 is shown in figure \ref{fig:orderhist}. Even though dependence of the residuals on filter order seems modest at most frequencies, it should be emphasized that even a factor 1.5 in residual noise can be significant especially with respect to NN, which directly contributes to the detection band. These results suggest that an adaptive filter technology that includes modifications of the filter order could lead to significantly decreased residuals.

\begin{figure}[t]
 \centerline{\includegraphics[width=0.7\textwidth]{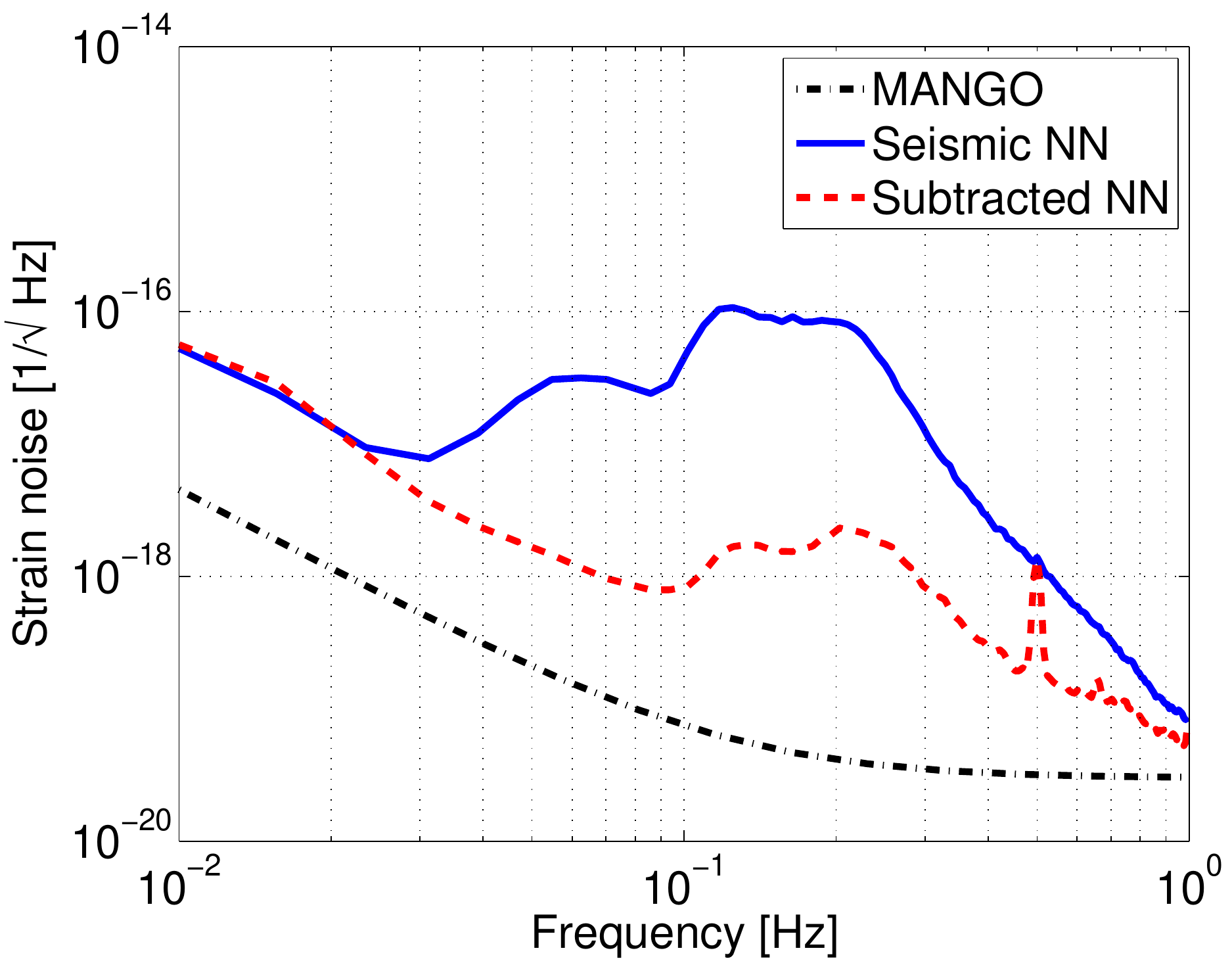}}
 \caption{The MANGO GW detector target spectrum, as well as the NN estimate and its residual after suppression by the factor that was achieved with seismic data. There is about a factor of 50 subtraction across the microseism. A further order of magnitude subtraction would be required to achieve MANGO GW detector target sensitivity.}
 \label{fig:MANGO}
\end{figure}

Next, we will outline the connection of our results to seismic NN subtraction. The subtraction of seismic NN will be more challenging than the demonstrated seismic-noise subtraction. Most importantly, it has been shown in previous simulations and theoretical work that reduction of NN from Rayleigh waves by a factor 50 or more would require a large number of seismometers \cite{BeEA2010,DHA2012,HaEA2013}, which follows from properties of the Rayleigh-wave field and the way it produces NN. The optimal design of the seismometer array will be a major challenge. There are also similarities between Newtonian and seismic-noise subtraction. Their performance can be limited by weaker contributions from other wave types such as compressional and shear waves generated directly by seismic sources, or by seismic scattering. The effect is similar for seismic and Newtonian-noise subtraction, since it is the spatial two-point correlation of the seismic field that determines the impact of diverse wave-type composition on NN (see \cite{BeEA2010} for details), which is the same function that determines the performance of the Wiener filter used for seismic-noise subtraction. Therefore, in this respect, our subtraction results are representative not only for seismic-noise cancellation, but also for NN cancellation. One can conclude that, if technical issues as described below are neglected, a factor 50 NN reduction can be achieved if the maximal spacing between seismometers in horizontal direction is 300\,m and 500\,m in vertical direction. In other words, at the level of achieved subtraction residuals, contributions from other wave types should not impede NN subtraction performance for the given seismometer spacing. The only way that this conclusion could turn out to be wrong due to properties of the seismic field is that the geologic conditions local to the Homestake mine are significantly more favorable in terms of seismic scattering than at areas at some distance to the mine, which are however still close enough to contribute to seismic NN at a significant level. One should keep in mind though that a seismometer spacing of a few hundred meters is very small for these frequencies, and that this could easily amount to a total of a few thousand seismometers required for a factor 50 NN subtraction. Now, based on this discussion, it may well be possible that the achieved seismic-noise subtraction is already limited by the presence of other wave types, especially since only 2 reference channels were used. This needs to be tested in the future using a larger seismic array that would be able to disentangle contributions from different wave types and possibly lead to better subtraction performance. These experiments will also show if a factor 50 suppression or higher can be achieved with rarer seismometer spacing. Nonetheless, we venture to present our subtraction results as an equivalent reduction of NN produced by Rayleigh waves. Figure \ref{fig:MANGO} shows the MANGO GW detector target spectrum, as well as the NN estimate based on the model presented in \cite{HaEA2013} and its residual after coherent cancellation using the achieved seismic-noise reduction. Another factor 20 reduction would be missing for the MANGO noise target. Other technical differences between seismic and Newtonian-noise subtraction should be mentioned. For example, a large number of reference channels used to calculate a NN Wiener filter may well come with additional numerical challenges, since the dimension of the correlation matrix that needs to be inverted for the Wiener filter would be much larger. In addition, the design of a seismic array that can provide a factor 50 NN reduction at low frequencies needs to address the additional problem that the seismic noise monitored by each seismometer will produce gravity perturbations that are strongly correlated between different test masses, which are therefore partially rejected by the differential readout of the detector (note that this common-mode rejection is included in the NN model of figure \ref{fig:MANGO}). This further increases potential numerical problems of the NN subtraction scheme, since the Wiener filter needs to incorporate the same effect by combining reference channels in a way that common-mode signals are cancelled at its output, and still be able to measure the underlying correlation of its output with the GW channel of the detector. Therefore, it should be clear that our results, though directly relevant to NN subtraction, demonstrate an advance concerning only one of many problems.  

Before concluding this section, we want to discuss further details of the seismic NN model used here. First, building a 0.1\,Hz detector underground has no significant influence on seismic NN. Seismic waves have lengths of a few tens of kilometers at these frequencies, and therefore relevant properties of seismic fields and seismic NN do not change with detector depths (for feasible depths). Second, the seismic NN model makes simplifications as outlined in detail in \cite{HaEA2013}. The most important consequence of these simplifications such as flat surface and homogeneous ground is that seismic scattering is not included in the NN model. Whereas this is certainly not an accurate representation of reality, our results show that these effects do not matter at frequencies around 0.1\,Hz at a level that is a factor 50 below the observed seismic-noise (assuming that seismic scattering at the Homestake mine is representative for the entire region). Therefore, at this level of accuracy, our results can be taken as a validation of these approximations. We want to emphasize though that our findings here cannot be used to draw similar conclusions concerning NN subtraction at much higher or lower frequencies, and it is possible that the situation changes drastically when noise suppression at 0.1\,Hz beyond a factor 50 is tried with larger arrays. 

\section{Conclusion}
\label{sec:Conclusion}

In this paper, we have investigated limits of coherent seismic-noise subtraction, which serves as a first test bed for the more challenging problem of seismic NN subtraction. The main difference between the two problems is that a much larger number of seismometers will be required for seismic NN subtraction. The question of optimal array design and many technical issues to calculate Wiener filters based on a large number of reference channels are not addressed by our study. However, our results allow us to put constraints on the effect from seismic scattering on coherent NN subtraction with the conclusion that at least a factor 50 NN reduction should in principle be feasible at the Homestake site around 0.1\,Hz, provided that seismic scattering at the Homestake site is representative for seismic scattering of the entire region that needs to be included for NN estimates.

We have demonstrated that we can achieve more than an order of magnitude seismic-noise cancellation between about 0.05-0.5\,Hz using Wiener filters with only a few seismometers separated by a distance of order 500\,m. We have also shown that this subtraction performance can be achieved without regularly updating the filter, indicating that the average properties of seismic fields at Homestake do not change significantly over timescales of weeks in this frequency band. This is beneficial for realizations in future GW detectors as it simplifies the application of the method to their output. However, in the attempt to optimize noise cancellation, it was found that filter order plays an important role. At frequencies below 0.1\,Hz, subtraction residuals varied almost by an order of magnitude for filter orders between 1 -- 1000. Whereas continuous optimization of filter order may not be feasible in many applications, especially in system control, the results also show that there are ranges of filter order with near-optimal subtraction performance over a broader range of frequencies. These filter orders maintain near-optimal performance over days.

As shown in various publications in the past \cite{GiEA2003,DrEA2012,DeEA2012}, Wiener filters can be used to efficiently subtract noise from data off-line, or in real time by means of feed-forward noise cancellation. In this paper, we focused on subtraction of coherent seismic noise, but many other applications are conceivable and have already been applied in previous generations of GW detectors, often forming an important part of the detector design. Wiener filters can subtract broadband noise as well as narrow-band features such as noise lines, and they will therefore play an important role in future detectors, and also serve as a starting point for the development of more advanced filter technologies.

These results are important as we have achieved factor of 50 subtraction in the low-frequency GW detectors band. For seismic NN, you would need another factor of 20 reduction to achieve $10^{-20}$ in strain sensitivity. Our work corresponds to the first test bed of noise subtraction in this band for these detectors. Although there is no direct connection to LIGO-like interferometers due to the low-frequency band, we are exploring pushing Wiener-filter subtraction to its limits and how to maximize its efficacy. With the installation of a larger seismic array, hopefully with good data quality above 1\,Hz, subtraction above 1\,Hz can be explored. As of now, it is not possible to say whether we can expect NN subtraction at the same levels as achieved in our analysis. This will require a test of the ability to monitor the seismic wavefield with an expanded seismic array. 

Because the residual spectra also contain a microseismic peak, it is evident that noise cancellation is not only limited by instrumental noise. A theory that should be tested in the future is whether the residual peak is produced by body waves instead of surface waves. It is known that both wave types contribute to the microseisms, but it is not clear how this affects noise cancellation. Alternatively, it is possible that topographic scattering of seismic waves play a role. The Homestake seismic underground array will be expanded in the future to more seismometers. This will make it possible to carry out a number of important studies relevant to seismic-noise cancellation, which were impossible with the more limited array used here. The extended array will have the capabilities to distinguish between body and surface waves, which, as explained, will be important to explore and possibly understand seismic-noise cancellation limits. It remains to be tested whether a larger array with greater variation in station distances would yield even better subtraction over a broader range of frequencies, potentially down to the instrumental noise limit. This possibility also motivates the development of improved seismometers with reduced self noise performance.

\section{Acknowledgments}
MC was supported by the National Science Foundation Graduate Research Fellowship Program, under NSF grant number DGE 1144152.
VD is a member of the LIGO Laboratory, supported by funding from United States National Science Foundation.
NC's work was supported by NSF grant PHY-1204371.
VM was supported by NSF grants PHY-0939669 and PHY-1344265.
LIGO was constructed by the California Institute of Technology and Massachusetts Institute of Technology with funding from the National Science Foundation and operates under cooperative agreement PHY-0757058.
This paper has been assigned LIGO document number LIGO-P1400042.

\section*{References}
\bibliographystyle{iopart-num}
\bibliography{references}

\end{document}